\newif\ifshort
\shortfalse
\documentclass[10pt,twocolumn,preprintnumbers,amsmath,amssymb,nofootinbib,superscriptaddress,prd]{revtex4-1}
\usepackage{graphicx,longtable,mathrsfs,color,array}
\usepackage[hidelinks]{hyperref}
\usepackage[usenames,dvipsnames]{xcolor} 
\usepackage{amssymb,amsmath,mathtools,mathrsfs,slashed} 
\usepackage{epsfig,subfigure,placeins,float} 
\usepackage{booktabs,longtable,ctable,multirow} 
\usepackage{exscale,relsize} 
\usepackage[normalem]{ulem} 
\usepackage{enumerate}
\usepackage{times, mathptmx} 
\usepackage[utf8]{inputenc}
\usepackage{color}
\usepackage{hyperref}
\usepackage{graphicx}
\usepackage{color}

\newcommand{\be}{\begin{equation}}
\newcommand{\ee}{\end{equation}}
\newcommand{\beq}{\begin{equation}}
\newcommand{\eeq}{\end{equation}}
\newcommand{\bea}{\begin{eqnarray}}
\newcommand{\eea}{\end{eqnarray}}

\newcommand{\bee}{{\bf{e}}}

\newcommand{\Gal}{{\rm G}}
\newcommand{\Lgw}{\mathcal{L}_{\rm GW}}

\newcommand{\dd}{\text{d}}

\newcommand{\nn}{{\nonumber}}

\usepackage[stable]{footmisc}

\definecolor{orange}{rgb}{1,0.5,0}

\newcommand\bees{\begin{eqnarray}}
\newcommand\ees{\end{eqnarray}}
\usepackage{graphicx,graphics}
\usepackage{bm}
\usepackage{tabularx}
\allowdisplaybreaks[1]

\newcommand{\abbr}[1]{\ifshort \else #1 \fi}

\begin{document}

\title{Detecting the anisotropic astrophysical gravitational wave
  background in the presence of shot noise through cross-correlations}
\author{David Alonso}
\thanks{Author list is alphabetised.}
\email{david.alonso@physics.ox.ac.uk}
\affiliation{Astrophysics, University of Oxford, DWB, Keble Road, Oxford OX1 3RH, UK}
\author{Giulia Cusin}
\email{giulia.cusin@physics.ox.ac.uk}
\affiliation{Astrophysics, University of Oxford, DWB, Keble Road, Oxford OX1 3RH, UK}
\affiliation{Universit\'e de Gen\`eve, D\'epartement de Physique Th\'eorique and Centre for Astroparticle Physics,
24 quai Ernest-Ansermet, CH-1211 Gen\`eve 4, Switzerland}
\author{Pedro G. Ferreira}
\abbr{\email{pedro.ferreira@physics.ox.ac.uk}}
\affiliation{Astrophysics, University of Oxford, DWB, Keble Road, Oxford OX1 3RH, UK}
\author{Cyril Pitrou}
\abbr{\email{pitrou@iap.fr}}
\affiliation{Institut d'Astrophysique de Paris, CNRS UMR 7095, 98 bis Bd Arago, 75014 Paris, France.}
\date{Received \today; published -- 00, 0000}

\begin{abstract}
  The spatial and temporal discreteness of gravitational wave sources leads to shot noise that may, in some regimes, swamp any attempts at measuring the anisotropy of the  gravitational wave background. Cross-correlating a gravitational wave background map with a sufficiently dense galaxy survey can alleviate this issue, and potentially recover some of the underlying properties of the gravitational wave background. We quantify the shot noise level and we explicitly show that cross-correlating the gravitational wave background and a galaxy catalog improves the chances of a first detection of the background anisotropy with a gravitational wave observatory operating in the frequency range $(10\,{\rm Hz},100\,{\rm Hz})$, given sufficient sensitivity. 
\end{abstract}

\maketitle
\section{Introduction}\label{Intro}
  The detection of gravitational waves is now in full flow heralding a new era in gravitational physics. One next frontier is the measurement and characterization of the gravitational wave background, a smooth but structured bath of gravitational radiation which may have come from the primordial Universe, but also from the plethora of gravitational wave signals emitted by different astrophysical sources from the beginning of stellar activity until today~\cite{Maggiore:2018sht}. One hopes that a clean measurement of this background will shed light on the physics of the early universe as well as the astrophysical properties of astrophysical sources (e.g. population of compact binaries).

  The astrophysical gravitational wave background (AGWB), i.e. the background generated by gravitational events at late cosmic times, is quantifiable through its isotropic energy density level and through the spatial angular power spectrum encoding its anisotropy. Existing data already place bounds on both the isotropic and anisotropic components \citep[e.g. see][]{2019arXiv190302886T, Mingarelli:2013dsa,Taylor:2013esa, Gair:2014rwa, 2019arXiv190308844T}. A detection of the isotropic level could come as early as 2020 \cite{Abbott:2017xzg}. 

  The AGWB in the LIGO band is mostly generated by a superposition of discrete events -- binary mergers, cataclysmic gravitational events, etc -- and, as such can be thought of a sequence of random processes. One starts off with the spatial distribution of the underlying density field, which can be modelled as a continuous random field (e.g. a realization of a multivariate Gaussian distribution on sufficiently large scales). Sources of gravitational waves will arise from a discrete sampling of this underlying density field; the simplest approach is to assume that it is a spatial Poisson process where the variance is set by the local number density of GW sources which is a relatively complicated function of the underlying density field. Finally, the events that lead to gravitational waves will often also be discrete in time, leading to a third layer of stochasticity.

  The discrete nature of the processes underlying the AGWB lead to a source of noise which is familiar from the analysis of galaxy surveys -- {\it Poisson} or {\it shot} noise. If the discrete part of the process is sufficiently sparse (i.e. the number density of sources or the rate of gravitational wave events is sufficiently small) shot noise may dominate, making it impossible to characterise the anisotropy of the AGWB (i.e. the underlying smooth field). A clear derivation of the problem from first principles can be found in \cite{2019PhRvD.100f3508J}.

  It has been suggested that it could be possible to sidestep the shot noise problem by cross-correlating a GW map with a dense galaxy sample tracing the same large-scale structure \cite{Cusinnew}. In this situation, the shot noise level of the cross-spectrum is primarily driven by the density of the much denser galaxy survey (although the GW shot noise will still be a significant contribution to the signal to noise of the cross-correlation). In this paper we explore this claim and quantify how much one can alleviate the shot-noise problem in measurements of the anisotropy of the AGWB with current gravitational wave experiments. 

  This paper is structured as follows. In Section \ref{AAGWB} we present a succint synopsis of the anisotropy of the AGWB and of its cross-correlation with the galaxy distribution. In Section \ref{XCorr} we discuss shot noise (both spatial and pop corn) and we sketch the argument of why cross-correlating a map of the AGWB with a galaxy survey may mitigate the shot-noise problem. In Section \ref{Results} we compute signal-to-noise ratio (SNR) for auto-correlation and cross-correlations and we show that the SNR for cross-correlation is significantly larger than the auto-correlation one and has a mild dependence on the cut-off chosen to filter out resolvable GW sources. We also show that most of the SNR comes from $z<0.5$, and that the result depends very mildly on the number density of galaxies. The SNR is also enhanced when considering the likely higher rate of neutron star mergers on top of black hole mergers, hence this observable might be a realistic and promising target for present and future galaxy surveys.

We must clarify from the outset, that we only intend to explore the shot-noise problem here, in isolation from other sources of noise. Therefore, all results reported below must be understood as forecasts for a perfect experiment with no instrumental noise. They therefore represent the best-case scenario for the detectability of the AGWB in the presence of spatial and temporal shot noise.

\section{AGWB and galaxy angular power spectra} \label{AAGWB}
  The isotropic AGWB signal can be characterized by the energy density in gravity waves, $\rho_{\rm GW}$, per logarithmic frequency interval in units of the critical density, $\rho_c$, averaged over directions: $\bar \Omega_{\rm GW}\equiv\dd\rho_{\rm GW}/\dd\ln f/\rho_c$ where $f$ is frequency. It can also be written as the sum of contributions from sources located at all the (comoving) distances $r$ in the form $\bar \Omega_{\rm GW}(f)\equiv 4\pi \int \partial_r \bar \Omega_{\rm GW}(f,r) \dd r$. Each astrophysical model predicts a functional dependence for $\partial_r \bar \Omega_{\rm GW}(f,r)$ where we define 
  \be\label{link2}
    \partial_{r}\bar{\Omega}_{\rm GW}=\frac{f}{4\pi \rho_c} a^4\int \dd\Lgw\, \bar{n}_{\Gal}(\Lgw, r)\Lgw\,, 
  \ee
  $\bar{n}_{\Gal}(\Lgw, r)$ is the average physical number of galaxies at distance $r$ with gravitational wave luminosity $\Lgw$. We use here, for definiteness, the reference astrophysical model of \cite{Cusinnew}.

  Of interest in this paper is the anisotropy of the AGWB, which can be modelled as 
  \be\label{sssm}
    \delta{\Omega}_{\rm GW}(\bee)= \int \dd r\,\partial_{r} \bar{\Omega}_{\rm GW} {\delta}_{\Gal}(r\,\bee )\,,
  \ee
  when sub-leading contributions from peculiar velocities and metric perturbations are ignored~\cite{Pitrou:2019rjz}. The line of sight direction is given by the unit vector $\bee$, and we assume an inhomogeneous distribution of galaxies hosting the gravitational wave sources, characterized by a density $n_\Gal={\bar  n}_\Gal(r) (1+{\delta}_{\Gal}(\bf r))$, where $\bar{n}_\Gal(r) = \int \bar{n}_{\Gal}(\Lgw, r) \dd \Lgw$ is the mean density of galaxies in the Universe at comoving distance $r$.
 
  In the Limber approximation, the general expression of the angular power spectrum of the anisotropies, $C^{\rm GW}_\ell$ of the AGWB reduces to \citep{Cusin:2017fwz,Cusin:2018ump}
  \be\label{unique}
    C^{\rm GW}_\ell(f) \simeq \left(\ell+\tfrac{1}{2}\right)^{-1}\int \dd k\,P_\Gal(k) \left|\partial_r \bar  \Omega_{\rm GW} (f,r_\ell)\right|^2\,,
  \ee
  where $\ell$ is the multipole in the spherical harmonic expansion, $P_\Gal(k)$ is the galaxy power spectrum, and $r_\ell = (\ell+1/2)/k$. Thus a measurement of $C^{\rm GW}_\ell$ is sensitive to the shape of $\partial_r \bar\Omega$ and of $P_\Gal(k)$. For details and derivations see \cite{Cusin:2017fwz, Cusin:2018rsq, Cusin:2018ump, Cusinnew}. 

  Consider now a direct measurement of the galaxy distribution, and let us construct a weighted average of the galaxy overdensity of objects along the line of sight by~\footnote{We assume that all galaxies are observed, hence $W(r)$ is not a selection function but rather a weight used to combine the distance dependent overdensities $\delta_G(r)$. It can be considered as an artificial selection function $W(r)/[r^2 a^3 \bar n_\Gal(r)]$.}
  \be\label{EqDefDeltaG}
    {\Delta}^\Gal(\bee)=\int \dd r\,W(r){\delta}_{\Gal}(r \,\bee  )\,,
  \ee
  where the weight function $W(r)$ is normalized so $\int W(r) \dd r= 1$. The auto-correlation and cross-correlation with AGWB in the Limber approximation are given by 
  \be
    \label{cdelta}
    {C}^{\Delta}_{\ell} \simeq \left(\ell+\tfrac{1}{2}\right)^{-1}\int \dd k\,P_\Gal(k) \left|W(r_\ell)\right|^2\,,
  \ee
  and 
  \be
    \label{ccross}
    {C}^{{\rm GW},\Delta}_{\ell}(f) \simeq
    \left(\ell+\tfrac{1}{2}\right)^{-1}\int \dd k\,P_\Gal(k) \,W(r_\ell) \partial_r \bar  \Omega_{\rm GW} (f,r_\ell)\,. 
  \ee

  As we can see, and very much along the lines of what is done in large-scale structure studies \citep[e.g.][]{2016PhRvD..94h3517N,2017arXiv170609359K}, we have a full set of spectra and cross spectra, Eqns.~(\ref{unique}), (\ref{cdelta}) and (\ref{ccross}) which characterize the statistical properties of the data $[\delta {\Omega}_{\rm GW}({\bf e}), {\Delta}^\Gal({\bf e})]$. Measuring these spectra can give us a wealth of information about the underlying processes that lead to the generation of gravitational waves in the late Universe, see \cite{Cusinnew, Cusin:2019jhg}.

\section{Spatial and pop corn shot noise} \label{XCorr}
  In this section we derive how shot noise arises and how it affects auto and cross-correlations. We distinguish between the shot noise arising from the discreteness of gravitational wave sources in space and due to their Poisson nature in the time domain. This will be useful in our estimate of the SNR in the next section. We note that, while in the previous section, we have presented auto and cross-correlations in terms of angular power spectra, our discussion here will be in terms of real-space correlations.

\subsection{The shot noise between two Poisson processes} \label{XCorr.shot}
  Consider two discrete sets of points, $a$ and $b$. In a given pixel $p$ there are $N_p^{a,b}$ points of each type, of which $N_p^c$ are common to both sets. We will write the ensemble average of each quantity as $\langle N_p^x\rangle\equiv N^x$. In a given pixel, let us write $N_{p}^x=N^c_p+N^{x-c}_p$. Assuming Poisson statistics, the first two moments of the distribution are:
\begin{equation}
  \langle N^x_p\rangle = N^x,\hspace{12pt}
  \langle (N^x_p)^2\rangle - \langle N^x_p\rangle^2 = N^x.
\end{equation}
The covariance between $a$ and $b$ is therefore:
\begin{align}
 & \langle N^a_p N^b_p\rangle - \langle N^a_p\rangle\langle N^b_p\rangle\nonumber\\
  &= \langle (N^c_p)^2 + N^c_p\,N^{a-c}_p + N^c_p\,N^{b-c}_p + N^{a-c}_p\,N^{b-c}_p\rangle-N^aN^b\nonumber\\
  &\equiv (N^c)^2 + N^c + N^c\,N^{a-c} + N^c\,N^{b-c} + N^{a-c}\,N^{b-c}\nn\\
  &\quad-(N^c+N^{a-c})(N^c+N^{b-c})\nn\\
  &=N^c\,,
\end{align}
where in the second line we have used the fact that $N^c$, $N^{a-c}$ and $N^{b-c}$ are all uncorrelated. We thus see that the cross-variance of two Poisson samples is equal to the number of events in the intersections of the two samples, i.e. Cov($N^a_p, N^b_p$)=$N^c$. We will use this result in the next sections.

\subsection{AGWB-galaxy count cross-correlations and shot noise}\label{XCorr.xcorr}

  The gravitational wave density fluctuation $\delta{\Omega}_{{\rm GW},p}$ in a  pixel $p$ is given by the cumulative flux of all gravitational wave sources along the line of sight $p$. Let us discretize this line of sight into intervals of  comoving distance $r$. Discretizing also the range of GW luminosities, $\mathcal{L}_{\rm GW}$, and ignoring  metric perturbations and peculiar velocities, we can write
  \begin{equation}\label{disc_gw}
    {\Omega}_{{\rm GW},p}=\frac{f}{\rho_c\,\theta_p^2}\sum_r\sum_{\Lgw} N^{\Lgw}_{r,p}\frac{\Lgw}{4\pi(1+z)\,r^2}\,,
  \end{equation}
  where $N^{\Lgw}_{r,p}$ is the number of sources in pixel $p$, in the radial bin $r$ and in the luminosity bin $\Lgw$, and where $\theta_p^2$ is the area of the pixel. In the continuum limit, taking the ensemble average, and writing
  \begin{equation}
    \left\langle N^{\Lgw}_{r,p}\right\rangle=a^3\,r^2\theta_p^2\,\dd r\,\dd\Lgw\,\bar{n}_{\rm G}(\Lgw,r)\,,
  \end{equation}
  we find 
  \begin{equation}
    \left\langle {\Omega}_{{\rm GW},p}\right\rangle=\int \dd r\int \dd\Lgw \frac{f\,a^4\,\Lgw}{4\pi\,\rho_c}\bar{n}_\Gal(\Lgw,r)\,,
  \end{equation}
  to recover the integral over the radial coordinate of Eq.\,(\ref{link2}). 

  On the other hand, the weighted galaxy number per solid angle along pixel $p$, $\Delta^\Gal_p$, is simply given by
  \begin{equation}\label{weights}
    1+ \Delta^\Gal_p=\sum_r \frac{W(r)}{\theta_p^2 r^2 a^3 \bar n_\Gal(r)} N^\Gal_{r,p}\,,
  \end{equation}
  where $N^\Gal_{r,p}$ is the number of galaxies in pixel $(r,p)$,whose average is
  \begin{equation}
    \left\langle N^\Gal_{r,p}\right\rangle=a^3 r^2\theta_p^2\,\dd r\,\bar{n}_\Gal(r)\,,
  \end{equation}
  and by construction we have $\langle 1+ \Delta_p^\Gal \rangle = 1$ as $\sum_r W(r)\,\dd r=1$.

  Assuming purely Poisson statistics for both $N^\Gal_{r,p}$ and $N^{\Lgw}_{r,p}$, we can now compute the variance of the different auto- and cross-correlations~\cite{Canas-Herrera:2019npr}. 
  \\
  \noindent
  {\it AGWB auto-correlation:}
  \begin{align}\nonumber
    &\left\langle {\Omega}_{{\rm GW},p}{\Omega}_{{\rm GW},p'}\right\rangle-\left\langle {\Omega}_{{\rm GW},p}\right\rangle\left\langle{\Omega}_{{\rm GW},p'}\right\rangle\\\nonumber
    &\hspace{12pt}=\delta_{pp'}\sum_{\Lgw}\sum_r\left(\frac{f\,\Lgw}{4\pi(1+z)r^2\rho_c\,\theta_p^2}\right)^2\left\langle N^{\Lgw}_{r,p}\right\rangle\label{VarAuto}\\
    &\hspace{12pt}=\frac{\delta_{pp'}}{\theta_p^2}\int\frac{\dd r}{r^2}\,\frac{1}{a^3\bar{n}_\Gal} \left(\partial_r\bar{\Omega}_{\rm GW} \right)^2,
  \end{align}
  where, in the last line, we have taken the continuum limit and we have assumed that all galaxies have the same GW luminosity, i.e.  $\bar{n}_\Gal(\Lgw,r)=\delta(\Lgw-\Lgw^0)\bar{n}_\Gal(r)$.
  \\
  {\it Number counts auto-correlation:}
  \begin{align}\nonumber
    &\left\langle \Delta_p\Delta_{p'}\right\rangle-\left\langle \Delta_p\right\rangle\left\langle\Delta_{p'}\right\rangle\\\nonumber
    &\hspace{12pt}=\delta_{pp'}\sum_r \left[\frac{W(r)}{\theta_p^2 r^2 a^3 \bar n_\Gal(r)} \right]^2\left\langle N^\Gal_{r,p}\right\rangle\\
    &\hspace{12pt}=\frac{\delta_{pp'}}{\theta_p^2}\int\frac{\dd r}{r^2}\,\frac{1}{a^3\bar{n}_\Gal} W^2(r)\label{VarDelta}
  \end{align}
  \\
  \noindent
  {\it  AGWB - number counts cross-correlation:}
  \begin{align}\nonumber
    &\left\langle {\Omega}_{{\rm GW},p}\Delta_{p'}\right\rangle-\left\langle {\Omega}_{{\rm GW},p}\right\rangle\left\langle\Delta_{p'}\right\rangle\\
    &\hspace{12pt}=\frac{\delta_{pp'}}{\theta_p^2}\int\frac{\dd r}{r^2}\,\frac{1}{a^3\bar{n}_\Gal} W(r) \partial_r\bar{\Omega}_{\rm GW}\,,\label{VarCross}
  \end{align}
  where we have assumed a monochromatic GW luminosity function, and that all galaxies emit GWs (and therefore ${\rm Cov}(N^{\Lgw}_{p,r},N^\Gal_{p,r})= \langle N^{\Lgw}_{r,p}\rangle$ as shown in the previous sub-section \ref{XCorr.shot}).

  We observe that the integral in Eq.\,(\ref{VarAuto}) diverges at the lower limit, when $r=0$, hence the contribution of Poisson noise of the AGWB auto-correlation depends on the cut-off used to regularize it. The reason for this divergence is that, for fixed luminosities, the flux of nearby sources increases like $\sim r^{-2}$, and therefore the very few closest sources end up dominating the total GW intensity across the sky. From an observational point of view, the physical quantity which sets the cut-off  is the observed flux: sources with a flux above a given threshold can be resolved and filtered out of the data. Given that the flux per unit frequency $\Phi$ from a source in $z$ is related to the luminosity per unit frequency through $\Phi(f)=(1+z) \Lgw/(4\pi d_L^2)$, we have that a lower bound on $\Phi$ is translated into a lower bound in redshift and an upper bound in luminosity\footnote{More precisely, it defines the region of integration in the plane $(z, \Lgw)$.}. Assuming that all galaxies have the same associated luminosity, the cut-off on flux directly translates into a lower cut-off in redshift (or analogously in $r$).

\subsection{Pop-corn shot noise}\label{pop}
  So far we have only considered the effect of the spatial discreteness of the sources of gravitational waves. In the frequency band of terrestrial interferometers, e.g. the LIGO-Virgo frequency band, the dominant contribution to the background comes from the merging phase of the evolution of  solar-mass compact objects. The signal is ``pop-corn''-like: events are separated in time and with almost no temporal overlap. Thus, there is a second shot-noise component due to the fact that events are discrete in time, and only some of them will contribute to the GW intensity mapped in a given time period. In this paper we focus on the contribution to the background coming from mergers of binary black hole systems.

  To compute this pop-corn shot noise we need to use the fact that the number of galaxies is a Poisson variable and each galaxy has a given (small) probability $\beta_T$ of containing a merger during the observation time $T$, with a Poisson distribution. We then use properties of compound Poisson distributions, see e.g. Ref.~\cite{2019PhRvD.100f3508J}. The only difference brought by this pop-corn noise on the results of \ref{XCorr.xcorr} is that the variance of the AGWB auto-correlation gets a correction prefactor of the form $(1+1/\beta_T)$, but the variance of cross-correlation (and galaxy auto-correlation) remain the same~\footnote{Let us denote $N^{\rm GW} =\sum_i^N y_i$ the total number of GW events in a pixel. $N$ is the number of galaxies in that same pixel which follows a Poisson distribution of average $\langle N \rangle$, and the $y_i$ also follow a Poisson statistics of average $\langle y_i \rangle =\beta_T$ due to the pop-corn nature of GW events. The compound statistics is found by averaging first over the statistics of the $y_i$ at fixed $N$ and then over the statistics of $N$. One finds easily ${\rm Cov}(N^{\rm GW},N^{\rm GW}) = \langle N \rangle (\beta_T + \beta_T^2)$ and ${\rm Cov}(N^{\rm GW},N) = \langle N \rangle \beta_T$. To compute the modification brought by the pop-corn nature (due to small values of $\beta_T$) we must form the ratio of these expressions with their asymptotic behaviour when $\beta_T \to \infty$. Hence we find for the auto-correlation of $N^{\rm GW}$ a modification factor $(\beta_T + \beta_T^2)/\beta_T^2=1+\beta_T^{-1}$, whereas the modification factor for cross-correlation is trivially $\beta_T/\beta_T = 1$.}. 

  The value of $\beta_T$  can be estimated
  \be\label{beta}
    \beta_T=\frac{T}{a^3 \bar{n}_\Gal}\times\frac{\dd\mathcal{N}}{\dd t \dd V}\,,
  \ee
  where $\dd\mathcal{N}/\dd V/\dd t$ is the merger rate per units of observed time and volume. Consider the upper bound for the merger rate \cite{2018arXiv181112907T}\footnote{In \cite{Fishbach:2018edt} it is also found that the inferred merger rate is consistent (at the $68\%$  confidence level) with being uniform in a comoving volume and source frame time.}
  \be\label{rate}
    \frac{\dd\mathcal{N}}{\dd t \dd V} <\frac{\dd\mathcal{N}}{\dd t_m \dd V} \sim 100\,\text{Gpc}^{-3}\text{yr}^{-1}\,,
  \ee
  where $t_m$ is the comoving time of the source, i.e. $t=(1+z)\,t_m>t_m$. Then using a constant comoving galaxy density $a^3 \bar{n}_{\Gal}\sim 0.1$ Mpc$^{-3}$, we find $\beta_T/T< 10^{-6}/{\rm yr}$. It follows that in the Hz band the shot noise of the AGWB auto-correlation (dominated by pop corn shot noise) is enhanced typically by a factor $10^{6}$ with respect to the shot noise in the mHz band (which is purely of spatial type). On the other hand, the shot-noise level of the cross-correlation stays the same over the whole frequency range and no enhancement due to the stochasticity in time of sources is present. This highlights the power of cross-correlation. We stress that this is just an order of magnitude estimate. To derive more accurate predictions for SNR in the next section, we will need to keep track of all the redshift factors in Eq.\,(\ref{beta}). 

\section{Results}
\label{Results}
\begin{figure*}[!htb]
  \includegraphics[width=0.49\linewidth]{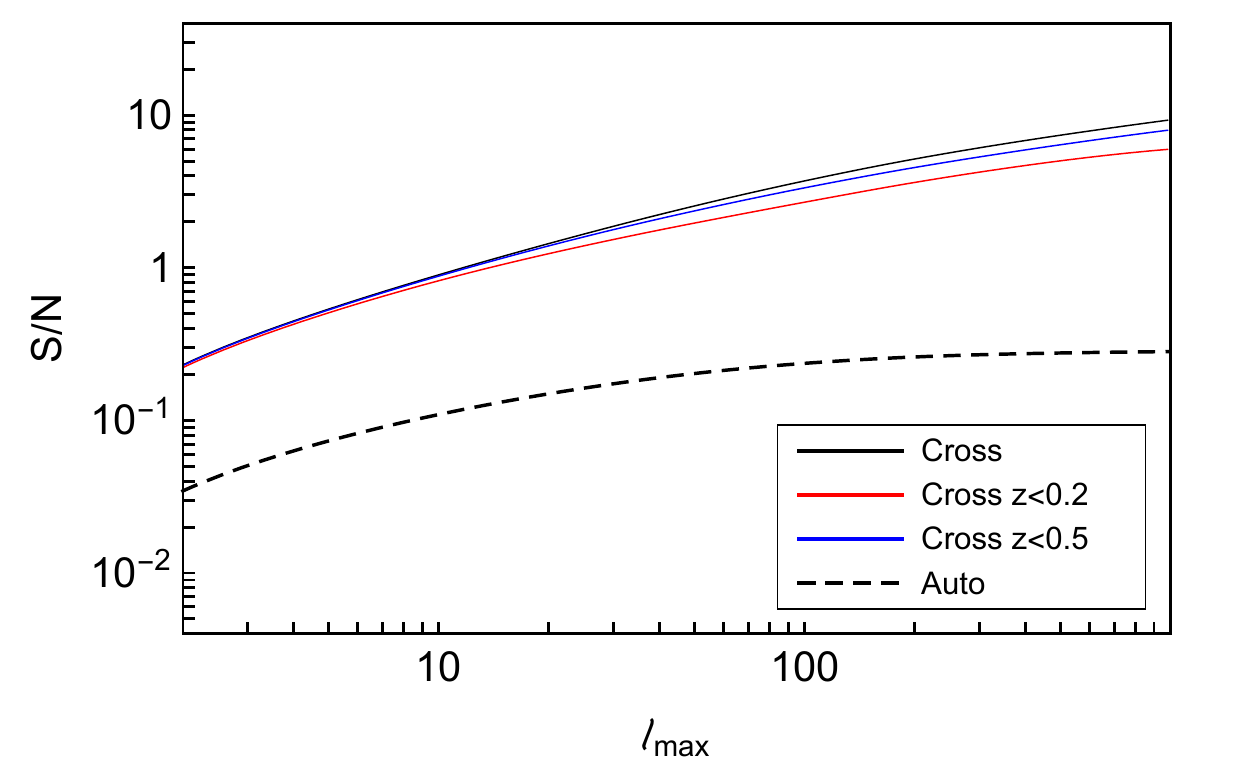}
  \includegraphics[width=0.49\linewidth]{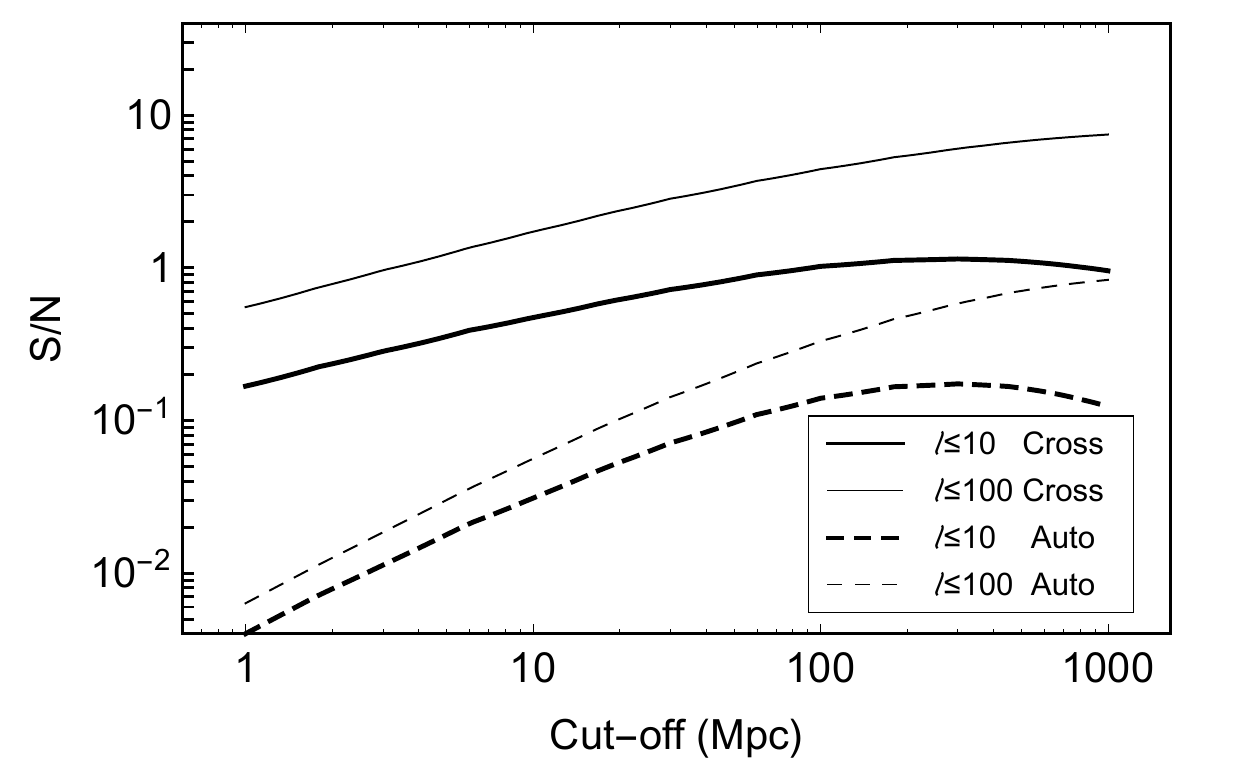} \\
  \includegraphics[width=0.49\linewidth]{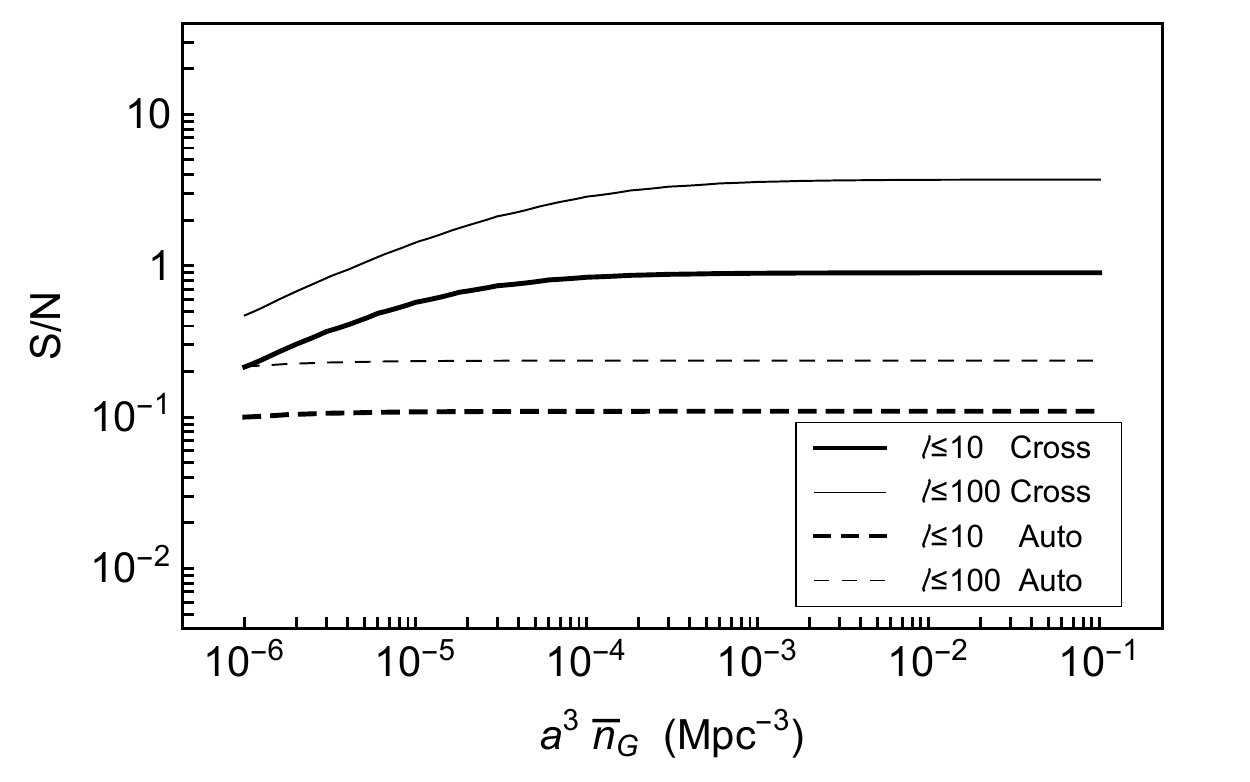} 
  \includegraphics[width=0.49\linewidth]{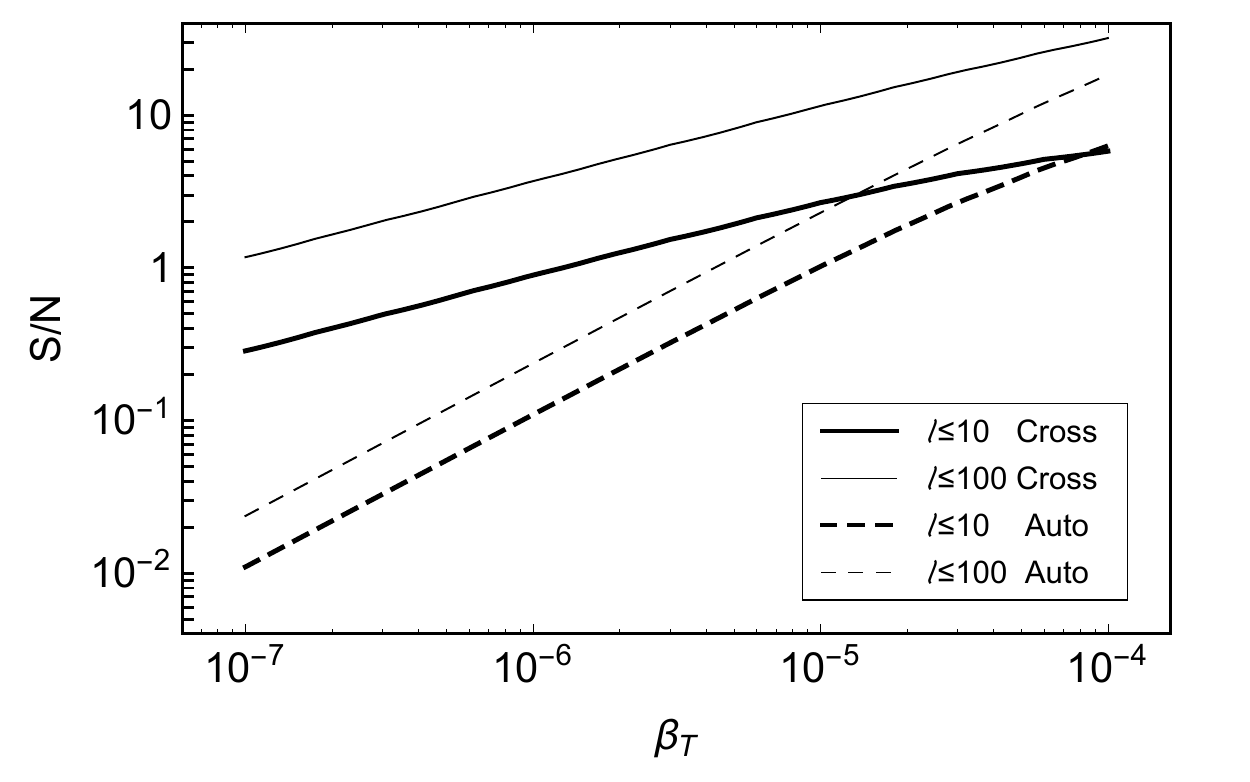} 
  \caption{\label{SNFig} {\it Top left :} cumulative SNR for AGWB auto-correlation (black dashed line) and cross-correlation (black cont. lines) with galaxy number counts using the optimal galaxy weight (and its restrictions to redshift bins in colors). {\it Top right :} dependence on the cut-off for two different maximum multipoles ($\ell_{\rm max}=10$ in thick lines and $\ell_{\rm max}=100$ in thin lines). Auto-correlations are in dashed lines, and cross-correlations with the optimal galaxy weight are in continuous lines. {\it Bottom left :} same curves varying instead the galaxy number density $a^3 \bar n_\Gal$. {\it Bottom right : } same curves varying instead the pop corn enhancement factor $\beta_T$. When not varied, the cut-off distance is 60 Mpc, the comoving galaxy density is $0.1\,{\rm Mpc}^{-3}$, and the pop corn enhancement factor is $\beta_T=10^{-6}$.}
\end{figure*}

  Before we embark on assessing the impact of cross-correlations, we note that the weight function, $W(r)$, should be chosen so as to maximize the SNR of cross-correlation. This can be done as long as radial information (i.e. accurate redshifts) are available for all galaxies in the survey we cross-correlate with, which we will assume here. As detailed in Appendix \ref{AppOptimal}, the optimal weights can be derived in terms of a Wiener filter, finding the result
  \be\label{Optimal}
    W(r) \equiv \frac{4\pi \partial_r \bar \Omega_{\rm GW}}{\bar \Omega_{\rm GW}}\,.
  \ee
  Physically this means that we approximately weight all galaxies by a $1/r^2$ factor, hence mimicking the properties of a background mapped in intensity. In principle this means that the spatial shot noises of auto and cross-correlations (Eqs. (\ref{VarAuto}-\ref{VarCross})) have exactly the same expressions (up to normalisation factors $4\pi/\bar \Omega_{\rm GW}$). In detail this is not exactly the case since the full expressions for galaxy numbers and for the GW background also involve sub dominant metric and velocity contributions, as well as the dominant galaxy overdensity term in eqs.\,(\ref{EqDefDeltaG}-\ref{ccross}). See \cite{Cusin:2017fwz,Cusin:2017mjm, Cusin:2018rsq, Cusin:2018ump, Cusinnew} for details. This implies that the optimal weight function found from the Wiener filter must differ slightly from \eqref{Optimal}.

  We can now estimate the SNR of the cross-correlation in the Hz (LIGO-Virgo) frequency band. We assume that shot noise is the only noise component, i.e. we assume an ideal experiment with no instrumental noise.  The SNR of the AGWB auto-correlation is given by~\footnote{The optimal full sky $C_\ell$ estimator for two observables $x$ and $y$ is $\hat  C^{xy}_\ell = \sum_{m} a^{x\star}_{\ell m} a^y_{\ell m}/(2\ell + 1)$. Its variance is easily deduced from the assumed Gaussianity of the $a^{x,y}_{\ell m}$, and it allows to deduce the SNR from a Fisher matrix analysis. The prefactor $2(2\ell+1)$ for the auto-correlation SNR (instead of the usual cosmic variance $(2\ell+1)/2$) is due to the fact that the signal, which is the amplitude of the GW background, appears quadratically in the observables (the $C_\ell$).} 
  \be\label{SN11}
    \left(\frac{S}{N}\right)_{{\rm GW}}^2 = \sum_\ell 2 (2\ell+1)\left(\frac{C_{\ell}^{{\rm GW}}}{C_{\ell}^{{\rm GW}}+N_{\ell}^{{\rm GW}}}\right)^2\,,
  \ee
  while the one of the cross-correlation is given by 
  \begin{align}\nonumber
    &\left(\frac{S}{N}\right)_{{\rm GW},\Delta}^2 \\
    & \,\,=\sum_\ell \frac{(2\ell+1)(C_{\ell}^{{\rm GW},\Delta})^2}{(C_{\ell}^{{\rm GW},\Delta}+N_{\ell}^{{\rm GW},\Delta})^2+(C_{\ell}^{{\rm GW}}+N_{\ell}^{{\rm GW}})(C_{\ell}^{\Delta}+N_{\ell}^{\Delta})}\,\,. \label{SN12}
  \end{align}
  The noise power spectrum $N_\ell$ is in fact given by the constants multiplying $\delta_{p p'}/\theta_p^2$ in Eqs.~(\ref{VarAuto}-\ref{VarCross}), as found from the discrete to continuous rule $\delta_{p p'}/\theta_p^2 \to \delta({\bf e}-{\bf e}')$. 

  In both (\ref{SN11}) and (\ref{SN12}) the dominant contribution to the denominator comes from the variance of AGWB auto-correlation $N_{\ell}^{\rm GW}$ due to the large pop-corn shot noise. Hence the SNR of cross-correlation will be typically enhanced with respect to the AGWB auto-correlation one. Having chosen the optimal weight \eqref{Optimal}, it is very easy to obtain analytic approximations. We first use that $C_\ell^{\Delta}=(4\pi/\bar\Omega_{\rm GW})C_\ell^{{\rm GW},\Delta}=(4\pi/\bar\Omega_{\rm GW})^2C_\ell^{\rm GW}$ (this is only approximate when including the subdominant metric contributions). Furthermore, we also find that all spatial shot noises are similarly related by factors $(4\pi/\bar\Omega_{\rm GW})$. Including the pop corn shot noise in GW auto-correlations, we then have $N_\ell^{\Delta} =(4\pi/\bar\Omega_{\rm GW}) N_\ell^{{\rm GW},\Delta} = (4\pi/\bar\Omega_{\rm GW})^2 N_\ell^{\rm GW}/(1+\beta_T^{-1})$.  For $\ell\gg1$ the $C_\ell$ scale roughly as $1/(\ell+1/2)$, as a consequence of the Limber expressions \eqref{unique}-\eqref{ccross} with large kernels. Using that $\beta_T\ll1$, the cumulative SNR of the auto-correlation scales as
  \be\label{SN11approx}
    \left(\frac{S}{N}\right)_{\rm GW}(\ell_{\rm max})\sim 2 \beta_T \alpha_{\rm cut} \sqrt{\ln \ell_{\rm max}}\,,
  \ee
  where we defined the cut-off dependent quantity $\alpha_{\rm cut}\equiv (\ell+1/2) C^{\Delta}_{\ell=1}/N^{\Delta}_\ell$, which is approximately constant for low $\ell$. A quick estimate for this coefficient is $\alpha_{\rm cut} \simeq a^3 \bar n_\Gal r^{\rm cut} \times \int P_\Gal(k)\dd k$, which is independent of the details of $\partial_r \bar \Omega_{\rm GW}$. For the SNR of the cross-correlation, one has
  \be\label{SN12approx}
    \left(\frac{S}{N}\right)_{{\rm GW},\Delta}(\ell_{\rm max})\sim \sqrt{2 \beta_T\alpha_{\rm cut} \ell_{\rm max} } \,,
  \ee
  where we used the scalings Eqs.~(\ref{cdelta}) and (\ref{ccross}). For an order of magnitude estimate, let us consider a cut-off at 60 Mpc for which $\alpha_{\rm cut}\sim 5\times10^4$. Then assuming integration time of one year and using for the value of $\beta_{T=1\rm  yr}$ its upper bound found in Sec.\,\ref{pop}, we have $\left(S/N\right)_{\rm GW}(\ell_{\rm max})\sim 0.1\sqrt{\ln \ell_{\rm max}}$ and $\left(S/N\right)_{{\rm GW},\Delta}(\ell_{\rm max}) \sim   \sqrt{0.1\,\ell_{\rm max}}$. The SNR up to a given $\ell_{\rm max}$, when using either the auto-correlation or the cross-correlations, is presented in the top left panel of Fig.~\ref{SNFig}. We observe that the behaviour with $\ell_{\rm max}$ is well described by the analytical scalings we have found. Note that our analysis differs significantly from Ref.~\cite{Canas-Herrera:2019npr}, where the constraints derived on posterior distributions are only cosmic variance limited.

  The dependence on the cut-off distance used when computing the GW pop corn shot noise is also illustrated in the top right panel. For these plots we have integrated the signal over the frequency range 10Hz$<f<$100Hz and assumed an integration time of 1 yr. We used the complete formula for background anisotropies of \cite{Cusin:2017fwz, Cusin:2017mjm}, where line of sight and velocity terms are added to the dominant galaxy clustering one. Given the rate of events \eqref{rate}, we expect an average of one black hole merger event in the sphere of radius $60\,{\rm Mpc}$ around us, for an observation time of ten years. Hence it is rather natural to choose a distance of that order as a cut-off, since this number might go up to the order of one GW event per year when including the likely higher rate of neutron star mergers.  

  With an instrument with extremely high sensitivity in the Hz band, GW mergers up to very high redshifts could be detected as individual events. These events could be filtered out and they would not act anymore as a noise component for the AGWB. The presence of a turning point in the top right panel of the figure, is due to the fact that low multipoles capture mainly contributions from low distance sources. As those sources are removed when increasing the cut-off distance, the signal of low multipoles is reduced while high multipoles are essentially unaffected. Simultaneously,  as we increase the cut-off distance, the noise gets reduced but the reduction is smeared over the whole multipole range since the $N_\ell$ are independent of $\ell$. This is why the turning point for the  total SNR up to $\ell_{\rm max}$ moves toward higher values cut-off distances as we increase $\ell_{\rm max}$. 

  In the bottom left panel we have shown the effect of reducing the number density of galaxies, and it is clear that the SNR is rather insensitive to its precise value as long as $a^3 \bar n_\Gal>10^{-4}$, which is comparable with current spectroscopic surveys.

  Finally, the dependence on the enhancement factor $\beta_T$ is illustrated in the bottom right panel. In this work we have studied only the contribution to the background coming from mergers of black holes in the Hz band. Another important background component in this band is given by merger of binary neutron star systems, see e.g. \cite{Cusinnew}. The merger rate of neutron stars is expected to be much higher (a factor 10) than the one of black holes~\cite{Abbott:2020uma}, the current upper limit for $\dd\mathcal{N}/\dd t/\dd V$  being $2810 \,\text{Gpc}^{-3}\text{yr}^{-1}$. Hence the pop corn shot noise will affect in a less severe way this background component as we expect $\beta_T$ to be typically larger by an order of magnitude.

  We find that most of the signal of the auto-correlation comes from low redshift ($z<0.5$). The Baryon Oscillation Spectroscopic Survey (BOSS, \citep{2015ApJS..219...12A}) has already covered wide swathes of the sky at these redshifts, and this coverage will keep improving in both number density and depth with future spectroscopic surveys, such as the Dark Energy Spectroscopic Instrument (DESI, \citep{2013arXiv1308.0847L}), the 4-metre Multi-Object Spectroscopic Telescope (4MOST, \citep{2019Msngr.175...50R}), and the Euclid satellite \citep{2011arXiv1110.3193L}. This indicates that  this is a realistic target for present galaxy surveys. Moreover, this also tells us that cross-correlating the AGWB with lensing  could be interesting observable if one wants to focus on the high-redshift GW signal, but it is not the best observable to look at to achieve a first detection of the anisotropies.

\section{Conclusion}
  The shot noise due the pop corn nature of GW sources in the Hz band is not a fundamental limitation that prevents one from getting information about the GW background anisotropies. Restricting to the background component coming from mergers of binary black hole systems, we have considered that shot noise and cosmic variance are the only noise components. In that idealized case, the SNR of the cross-correlation with a galaxy catalog is found to be much higher than that of the auto-correlation. The SNR of the cross-correlation is of order $\sim$10  for large ($\gtrsim100$) $\ell_{\rm max}$ for realistic galaxy number densities ($a^3 \bar n_{\rm G}\sim10^{-3}{\rm Mpc}^{-3}$). Moreover, we have shown that most of the signal comes from low redshift $z<0.5$, indicating that present galaxy catalogs can already be used to construct these cross-correlations.

While this analysis has shown that there is some promise in this method, it is useful to take a more conservative view of its feasibility with up and coming data, and what we may learn from such an observation. Currently, it is envisaged that all events out to approximately 1 Gpc will be resolved (of order $10^3$ in total). If we take this to be the effective cutoff we see that the SNR can be appreciable if we are able to resolve the map down to approximately $1^\circ$, i.e. $\ell_{\rm max}\sim 100$. But in this situation we need to face a few issues. For a start, we haven't included instrumental noise which will degrade the effective resolution and which rapidly leads to a degradation of the SNR. Furthermore, for this cut-off choice the signal will be primarily from GW sources beyond 1 Gpc or a $z\sim0.4$; the question then arises of how much information we can extract from the AGWB background about, for example, high redshift binary populations and merger rates, as compared to what might be inferred from the direct analysis of the resolved events at lower redshift.

  In this analysis we focused on the contribution of black hole mergers, but this study can be easily extended to the population of binary neutron stars. The neutron star merger rate being much larger than the one of black holes, this background component is expected to be less affected by shot noise than the black hole one studied in this work, as we expect a reduction of $\beta_T$ by one order of magnitude (i.e. $\beta_T \simeq 10^{-5}$ for one year of observation). Also, $\beta_T \propto T$, and therefore the SNR will keep on improving as more data is collected. The rate of improvement will $\propto \sqrt{T}$ and $\propto T$ for the cross-correlation and auto-correlation respectively. For very long total observation time (such that $\beta_T$ reaches $10^{-4}$), the significance of both observables becomes comparable.

  It is worth noting that, although it is always possible to avoid the offset in the AGWB power spectrum caused by shot noise, commonly called the ``noise bias'', by using only cross-correlation between different data splits \citep{Jenkins:2019nks} (a technique that is extensively used in the CMB community to avoid complicated instrumental noise biases), this will not mitigate in any way the impact of shot noise on the variance of the estimated power spectrum (which will be the dominant contribution). Cross-correlating with a denser sample that traces the same underlying structure, on the other hand, does lead to a significant mitigating factor (see Eqs. \ref{SN11} and \ref{SN12}). This has been used in large-scale structure surveys to study the clustering of sparse samples, such as damped Lyman-$\alpha$ systems \citep{2012JCAP...11..059F,2018JCAP...04..053A}.

Most importantly, we must emphasize the fact that our analysis has not accounted for any form of instrumental noise. For a realistic instrument, it is in principle not clear what the best strategy would be to carry out this cross correlation. One possibilities would be to use only resolved events. In this case the signal would be dominated by radiometer noise, and the detection would be limited by the small number of resolvable events. The second possibility would be to search for a background of unresolved events. In this case, the detection is likely to be limited by detector noise and the poor angular resolution of Earth-based facilities. This translates into an low effective $\ell_{\rm max}$, and from the top left panel of Figure \ref{SNFig} we infer that the associated SNR would probably remain below unity.  

LIGO-Virgo is expected to detect the isotropic component of the background with the design sensitivity  \cite{Abbott:2017xzg}. Given that the typical amplitude of the AGWB anisotropies are suppressed by a factor of at least $10^{-2}$ with respect to the monopole, this means that an improvement in design sensitivity of at least a factor 10 is necessary to get a detection (on the angular scales accessible given the diffraction pattern of the observatories). Einstein Telescope is expected to reach this sensitivity  threshold \cite{ETreport}. However, an improved sensitivity also implies that the catalogue of resolvable sources one can detect becomes much more complete and deep in redshift and one expects to have a broad redshift cover (up to $z\sim2-3$ at least) of all resolvable sources, with a much better angular resolution than the one of AGWB anisotropy.

A minor point is that in our analysis we have filtered out (from the background) the contribution from close sources. However, in a  realistic analysis the cut-off is actually in the received flux and not in distance. To refine our study, we should convolve our results with some distance distribution for a given brightness cut-off. However, since our signal to noise for the cross-correlation flattens out as a function of the cut-off, this will not  significantly alter the results.

  The situation described here is very different in the mHz band (e.g. the LISA band). In this case the background is composed by the superposition of signals from binary systems in the inspiralling phase. Since the duration of the inspiralling phase is much larger than typical observation times, the signals add up to form a continuous and almost stationary background. This is an \emph{intrinsic} (irreducible) background. The shot noise in the LISA band will therefore only be due to the discreteness in space of the GW sources, and will be a subdominant contribution to the total error budget (see also \cite{Cusin:2019jhg}). \\

  In summary, we have found that there are no intrinsic (i.e. shot-noise-like) noise components that constitute a fundamental barrier to obtaining information from the anisotropies in the AGWB in the Hz band, and that cross-correlating with a galaxy survey traceing the same underlying structures is a promising method to get a first detection of the anisotropies. This result holds in idealized case without instrumental noise, and a future work will be dedicated to applying this analysis to realistic GW detector networks and galaxy surveys. \\

  \textit{Acknowledgements ---} We are extremely grateful for discussions with Irina Dvorkin about the AGWB modelling. We would also like to thank Stefano Camera and Andrew Matas for useful insights and Carlo Contaldi and Niel Cornish for a careful reading of this manuscript. Some of the power spectrum calculations were carried out using the Core Cosmology Library (CCL) \citep{2019ApJS..242....2C}. DA acknowledges support from Science and Technology Facilities Council through an Ernest Rutherford Fellowship, grant reference ST/P004474/1. PGF and DA acknowledge support from the  Beecroft Trust. This  project  has  received  funding  from  the European  Research  Council  (ERC)  under  the  European  Union's Horizon  2020  research  and  innovation  programme  (grant  agreement No 693024)  and from the Swiss National Science Foundation.

\bibliographystyle{apsrev4-1}
\bibliography{Xrefs}


\appendix

\section{Optimal weights}\label{AppOptimal}
Consider a vector of $N$ measurements ${\bf x}=(x_1,...,x_N)$ that we want to linearly combine to find the best estimate of a given quantity $y$. Assuming Gaussian statistics, the probability for $y$ given ${\bf x}$ is
\begin{equation}
  -2\log p(y|{\bf x})={\bf z}^T{\sf C}_{zz}^{-1}{\bf z}-{\bf x}^T{\sf C}_{xx}^{-1}{\bf x},
\end{equation}
where we have defined the vector ${\bf z}=(y,x_1,...,x_N)$, and ${\sf C}_{ab}$ is the covariance matrix between ${\bf a}$ and ${\bf b}$ (${\sf C}_{ab}\equiv\left\langle{\bf a}{\bf b}^T\right\rangle$). Therefore ${\sf C}_{zz}$ is
\begin{equation}
 {\sf C}_{zz}=\left(
 \begin{array}{cc}
  C_{yy} & {\bf C}_{xy}^T \\
  {\bf C}_{xy} & {\sf C}_{xx}
 \end{array}
 \right)\,.
\end{equation}
The maximum-likelihood estimator for $y$ can be found by solving the equation $-2\partial_y\log p(y|{\bf x})=0$. After a little algebra, this estimator is
\begin{equation}
 \hat{y} = {\bf w}^T{\bf x}\equiv{\bf C}_{xy}^T{\sf C}_{xx}^{-1}{\bf x}.
\end{equation}
The linear coefficients ${\bf w}$ are the so-called \emph{Wiener filter}.

In our case, $y$ is the gravitational wave background in a given pixel $\Omega_{{\rm GW},p}$, and ${\bf x}$ is a vector of number count measurements along that pixel's line of sight $N_{r,p}$. Assuming Poisson error bars, the different covariance elements are
\begin{align}
 C_{r r'} \equiv {\rm Cov}(N_{r,p},N_{r',p})&=\delta_{rr'}\,\dd r\,r^2\,a^3 \bar{n}_\Gal(r)\,\theta_p^2\,,\\
 C_{r \Omega} \equiv  {\rm Cov}(N_{r,p},\Omega_{{\rm GW},p})&=\dd r\partial_r\bar{\Omega}_{\rm GW}\,.
\end{align}
Therefore the Wiener filter weights of the sum \eqref{weights} are
\begin{equation}\label{WienerWeight}
 \frac{W(r)}{r^2 a^3 \bar n_\Gal(r)} \propto \frac{\partial_r\bar{\Omega}_{\rm GW}}{r^2 a^3 \bar n_\Gal(r)}\,.
\end{equation}
Using these weights allows to build the most likely GW background in a
given pixel only from its galaxy number measurement binned in redshifts. Hence it is the good
variable to use for cross-correlating with the directly measured GW background of that pixel so as to constrain its global amplitude. Once properly normalized, the weights \eqref{WienerWeight} lead to the weight function \eqref{Optimal}.

Equivalently, the weights can be obtained from a least-squares analysis of the  variables $N_{r,p}\Omega_{\rm GW,p}$. This allows to build an estimator for the GW amplitude $A_p = 4\pi \Omega_{{\rm GW},p}/\bar \Omega_{\rm GW}$ as
\be
\hat{A}_p  \equiv \frac{\sum_r \left(\sum_{r'} C_{r' \Omega} {\cal C}^{-1}_{r'
      r}\right) N_{r,p} \Omega_{{\rm GW},p}}{\sum_{r r'} C_{r' \Omega } {\cal C}^{-1}_{r'
      r}  C_{r \Omega}}\,,
  \ee
where ${\cal C}_{r r'} \equiv {\rm Cov}(N_{r,p}\Omega_{{\rm GW},p}
\,,N_{r',p}\,\Omega_{{\rm GW},p}) =C_{r r'} C_{\Omega \Omega} + C_{r
  \Omega} C_{r' \Omega}$. Assuming that this covariance is
dominated by the pop-corn noise of the GW background, we can then
approximate ${\cal C}_{r r'} \simeq C_{r r'} C_{\Omega \Omega}$, from
which we infer again that the optimal weights are $\propto \sum_{r'} C_{r' \Omega}C^{-1}_{r' r}$, therefore recovering Eq.~\eqref{WienerWeight}.
\end{document}